\documentclass[10pt,twocolumn,letterpaper,superscriptaddress,prb,showpacs]{revtex4}

\usepackage{graphicx,amsmath,color,amssymb}
\usepackage[latin1]{inputenc}
\usepackage[sort&compress]{natbib}
\usepackage{epstopdf}
\usepackage{booktabs}
\usepackage[greek,english]{babel}
\bibpunct{}{}{,}{s}{}{,}

\usepackage[dvips]{epsfig}
\usepackage{subfigure,psfrag}
\usepackage{lscape}
\usepackage{varioref}
\usepackage{float}
\usepackage{tabularx}
\usepackage[version-1-compatibility]{siunitx}
\usepackage[normalem]{ulem}

\definecolor{darkblue}{rgb}{0,0,.4}
\definecolor{darkred}{rgb}{0.6,0,0}
      
\newcommand{\abs}[1]{\left|#1\right|}
\newcommand{\ket}[1]{\left| #1 \right>}
\newcommand{\bra}[1]{\left< #1 \right|}

\newcommand{\AlGaAs}{Al$_{0.3}$Ga$_{0.7}$As}
\newcommand\figref[1]{Fig.~\ref{#1}}
\renewcommand\eqref[1]{Eq.~(\ref{#1})}
\newcommand\secref[1]{Sec.~\ref{#1}}
\newcommand\DE{droplet epitaxy}
\newcommand{\OS}{oscillator strength}
\newcommand{\QE}{quantum efficiency}

\usepackage[bookmarksopen=true,
citecolor=blue,
colorlinks=true,
linkcolor=blue]{hyperref}

\begin{document}

\title{Decay dynamics and exciton localization in large GaAs quantum dots grown by droplet epitaxy}

\author{P.~Tighineanu}
\email{petru.tighineanu@nbi.ku.dk}
\affiliation{Niels Bohr Institute,\ University of Copenhagen,\ Blegdamsvej 17,\ DK-2100 Copenhagen,\ Denmark}
\author{R.~Daveau}
\affiliation{Niels Bohr Institute,\ University of Copenhagen,\ Blegdamsvej 17,\ DK-2100 Copenhagen,\ Denmark}

\author{E.~H.~Lee}
\affiliation{Center for Opto-Electronic Convergence Systems,\ Korea Institute of Science and Technology,\ Seoul,\ 136-791,\ Korea}
\author{J.~D.~Song}
\affiliation{Center for Opto-Electronic Convergence Systems,\ Korea Institute of Science and Technology,\ Seoul,\ 136-791,\ Korea}

\author{S.~Stobbe}
\affiliation{Niels Bohr Institute,\ University of Copenhagen,\ Blegdamsvej 17,\ DK-2100 Copenhagen,\ Denmark}
\author{P. Lodahl}
\email{lodahl@nbi.ku.dk}
\homepage{http://www.quantum-photonics.dk/}
\affiliation{Niels Bohr Institute,\ University of Copenhagen,\ Blegdamsvej 17,\ DK-2100 Copenhagen,\ Denmark}

\date{\today}

\small

\begin{abstract}
We investigate the optical emission and decay dynamics of excitons confined in large strain-free GaAs quantum dots grown by droplet epitaxy. From time-resolved measurements combined with a theoretical model we show that droplet-epitaxy quantum dots have a quantum efficiency of about \SI{75}{\percent} and an oscillator strength between 8 and 10. The quantum dots are found to be fully described by a model for strongly-confined excitons, in contrast to the theoretical prediction that excitons in large quantum dots exhibit the so-called giant oscillator strength. We attribute these findings to localized ground-state excitons in potential minima created by material intermixing during growth.
We provide further evidence for the strong-confinement regime of excitons by extracting the size of electron and hole wavefunctions from the phonon-broadened photoluminescence spectra. Furthermore, we explore the temperature dependence of the decay dynamics and, for some quantum dots, observe a pronounced reduction in the effective transition strength with temperature. We quantify and explain these effects as being an intrinsic property of large quantum dots owing to thermal excitation of the ground-state exciton. Our results provide a detailed understanding of the optical properties of large quantum dots in general, and of quantum dots grown by droplet epitaxy in particular.
\end{abstract}


\maketitle

\section{Introduction}
Semiconductor quantum dots (QDs) have attracted considerable interest over the past years as mesoscopic light emitters for all solid-state quantum electrodynamics experiments.\cite{reithmaier04,yoshie04,peter05,englund05,badolato05,hennessy07,lund08} Commonly denoted artificial atoms due to their discrete energy levels, QDs offer an attractive platform for tailoring light-matter interaction at the nanoscale for use in 
quantum-information protocols.\cite{shields07,obrien09,ladd10,lodahl12} A present challenge constitutes the realization of efficient single-photon sources, which generate a large number of flying qubits per unit time. Combined with their sub-poissonian statistics and large oscillator strength, highly efficient QDs are excellent candidates for use both in the Purcell regime as solid-state light-emitting diodes or single-QD lasers,\cite{michler04} and in the strong-coupling regime\cite{McKeever03} in exploring the realm of quantum optics on a solid-state platform.\cite{reithmaier04,peter05}

According to Fermi's Golden Rule, efficient emitters can be realized by enhancing the interaction strength between QDs and light. In the dipole approximation, the figure of merit of light-matter interaction is given by the oscillator strength of the emitter times the projected local density of optical states (LDOS) at the position of the emitter.\cite{novotny06} The \OS\ is a dimensionless quantity defined as the ratio between the radiative decay rate of the emitter in a homogeneous medium and the decay rate of a harmonic oscillator, and is solely a property of the emitter. On the other hand, the LDOS quantifies the available density of optical states and is determined by the nanophotonic environment surrounding the QD. One way of enhancing light-matter interaction is to increase the LDOS by embedding the QD in an optical cavity or waveguide. For a cavity for example, the enhancement magnitude is determined by its quality factor and inverse mode volume. To increase light-matter coupling even further, it becomes imperative to engineer QDs with an enhanced \OS. In fact, it has been pointed out by Hanamura\cite{hanamura88} that for excitons in the weak confinement regime, that is, when the spatial extent of the confinement potential is larger than the exciton Bohr radius, the \OS\ is proportional to the QD-volume and is predicted to be orders of magnitude larger than that of strongly confined excitons (the so-called `giant oscillator strength' regime).\cite{stobbe12} 

Engineering QDs with giant \OS\ is not straightforwardly accomplished in practice because the QDs must have a uniform potential profile over length scales larger than the exciton Bohr radius. For instance, the commonly employed  In(Ga)As/GaAs QDs  suffer from inhomogeneous strain and alloy composition, which create localized potential minima thereby impeding the coherent distribution of the ground-state exciton over length scales comparable to the measured physical size of the QDs. Another fundamental issue is the large value of the exciton Bohr radius, which attains \SI{48}{\nano\meter} in InAs, as compared to only \SI{11}{\nano\meter} in GaAs.\cite{cardona10} Employing a modified LDOS near a semiconductor-air interface revealed a small \OS\ of large In(Ga)As/GaAs QDs corresponding to strongly-confined charge carriers.\cite{stobbe10} Interface-fluctuation GaAs QDs constitute another example of widely studied large semiconductor QDs\cite{gammon96} where indications of a giant oscillator strength have been reported.\cite{senellart05} 

\begin{figure*}[t!]
	\includegraphics[height=0.25\textheight]{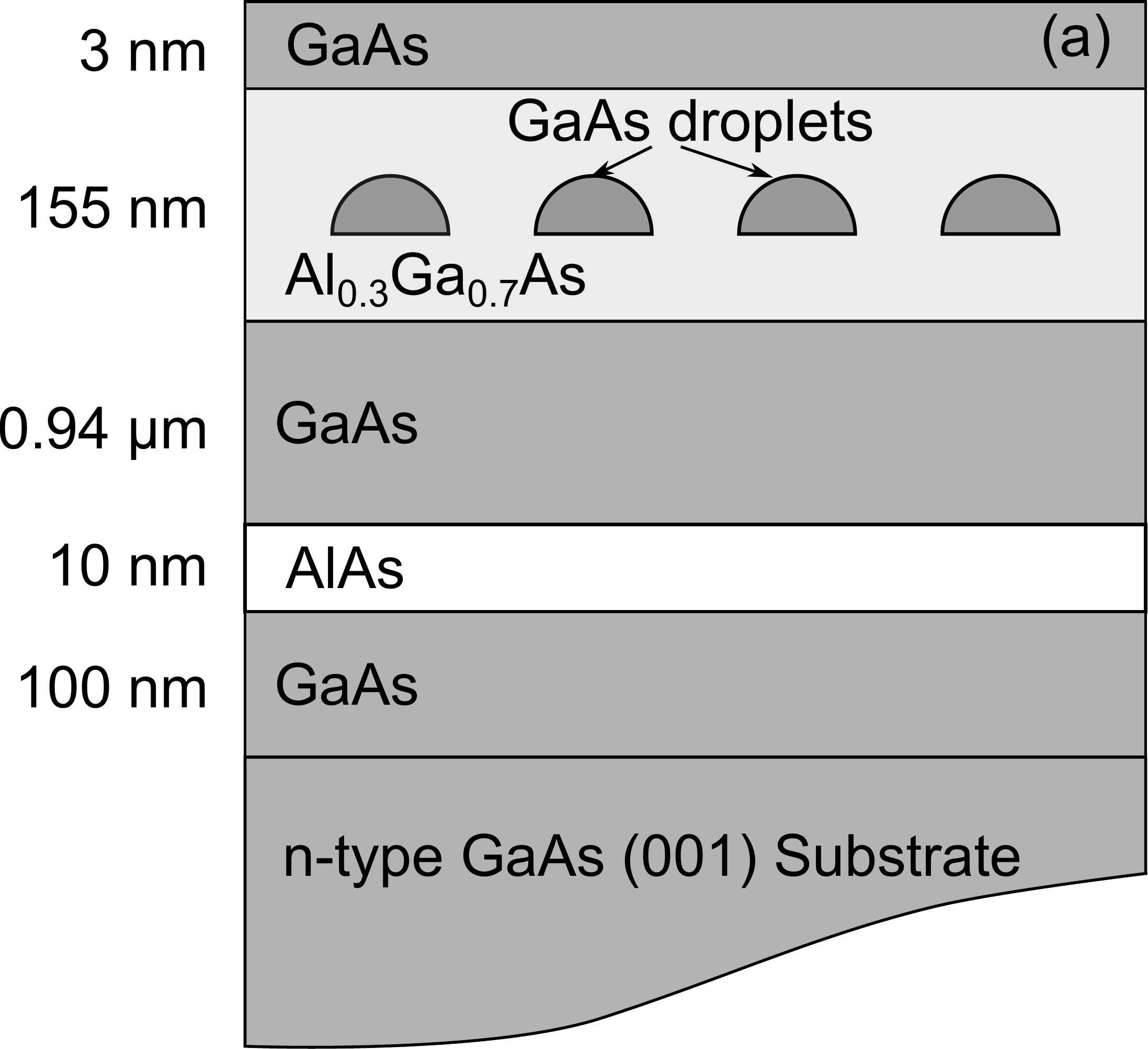}~\hspace{0.5cm}
	\includegraphics[height=0.25\textheight]{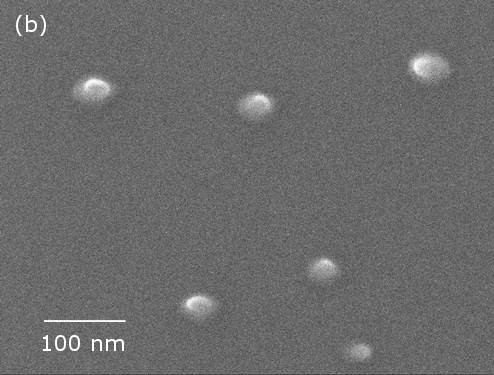}
	\caption{\label{fig:sampleProfile} (a)~Schematic of the cutaway profile of our sample (not to scale). (b)~Scanning electron micrograph data of the uncapped reference sample.}
\end{figure*} 

Droplet epitaxy\cite{koguchi93,watanabe00,watanabe01,mantovani04,wang06,ha11} is a powerful emerging growth technique, which is potentially capable of providing a solution to the aforementioned challenges. Recently, it has been shown that with a proper interplay among growth parameters, one can obtain droplet-epitaxy QDs with an optical quality (i.e., narrow linewidths) approaching that of Stranski-Krastanov (SK) InAs QDs.\cite{mano09} Self-assembled GaAs QDs grown by \DE\ are strain free because they are embedded in lattice-matched AlGaAs barriers,\cite{koguchi93} which renders two important advantages. First, there is no strain energy stored in the QDs, which would degrade the homogeneity of the potential profile, and second, strain-related structural defects are avoided. In this regard, physically large QDs can be realized, which could potentially exhibit strongly enhanced coupling to light. Moreover, \DE\ permits the growth of QDs with a very low surface density (a few QDs per \si{\micro\meter^2}), which enables their individual control and manipulation. Finally, QDs grown by \DE\ are promising for use in the visible spectrum where Si-based detectors attain maximum efficiency. 

Despite these important advantages, \DE\ is a relatively new technology and the droplet-epitaxy QDs lack detailed and systematic studies of their optical properties. In particular, their \OS\ and quantum efficiency have not been studied. From a physical point of view, the \QE\ is the probability that an exciton captured by the QD recombines radiatively. Being part of a solid-state system, QDs are prone to growth imperfections and, hence, to nonradiative decay channels, e.g., via carrier trapping by QD surface states.\cite{stobbe09} Unfortunately, little attention is being drawn in the literature to nonradiative decay and it is often implicitly assumed that the QDs decay purely radiatively. Nonradiative processes degrade the ability of QDs to generate single photons on demand, which is an important goal in the field of quantum-information processing.  Nonradiative rates have been difficult to quantify, since the measured recombination rate of the exciton is given by the sum of radiative and nonradiative components, and hence their individual contribution are not separated straightforwardly. By controllably modifying the LDOS, it has been shown recently that SK In(Ga)As QDs possess non-negligible nonradiative contribution with \QE\ between \SI{80}{\percent} and \SI{95}{\percent}.\cite{johansen08} Furthermore, large In(Ga)As QDs were found to exhibit a \QE\ of only 30 to \SI{60}{\percent}.\cite{stobbe10}

In this paper, we perform for the first time a systematic study of the decay dynamics of large QDs grown by \DE\ and measure the \OS\ and \QE . We present a detailed analysis of three individual QDs. Surprisingly, the \OS\ reveals that the excitons are in the strong-confinement regime despite the large size of droplet-epitaxy QDs. The small exciton size is cross-checked by quantitatively analyzing the phonon-broadened spectra. Our results are in qualitative agreement with the work of Rol et al.\cite{rol07} for GaN QDs, where a similar analysis revealed that the excitons are smaller than the QD size. The extracted \QE\ (70 to \SI{80}{\percent}) turns out to be lower than that of small In(Ga)As QDs\cite{johansen08} yet larger than the \QE\ of large In(Ga)As QDs.\cite{stobbe10} Our work confirms that nonradiative processes in semiconductor QDs have a profound impact on their optical properties. We also show that some QDs exhibit a pronounced reduction in their effective transition strength and quantum efficiency with temperature, which we attribute to coupling to excited states of the QD.

The paper is organized as follows. In \secref{sec:ExpProc} we outline the growth recipe of our sample and the experimental procedure in the optical experiments. \secref{sec:OptEmission} presents the spectral behavior of the photoluminescence (PL) while \secref{sec:OSQE} is devoted to analysing the decay dynamics of the excitons captured by the QDs, which enables extracting the \OS\ and \QE . \secref{sec:TOS} explores the temperature dependence of the excitonic decay dynamics. Finally, in \secref{sec:LOphonons} the phonon-broadened spectra are quantitatively analyzed.

\section{Sample Growth and Experimental Procedure}
\label{sec:ExpProc}

The sample used in our experiment was grown on an n-type GaAs (001) wafer. After thermal removal of surface oxides, \SI{0.1}{\micro\meter} GaAs, \SI{10}{\nano\meter} AlAs, \SI{0.94}{\micro\meter} GaAs, and \SI{50}{\nano\meter} \AlGaAs\ layers was grown successively at \SI{580}{\celsius}. Thereafter, GaAs QDs were grown by droplet epitaxy according to the following procedure. At a substrate temperature of \SI{300}{\celsius}, Ga atoms were injected onto the surface at a vacuum level of \SI{e-10}{\torr}. The amount of Ga is equivalent to the Ga content in two GaAs monolayers. After injection of As and subsequent crystallization, a \SI{20}{\nano\meter}-thick \AlGaAs\ layer was grown by migration-enhanced epitaxy, a technique used for growing high-quality heterointerfaces at low temperatures.\cite{horikoshi88} The temperature was then raised back to \SI{580}{\celsius} and an additional \SI{85}{\nano\meter} \AlGaAs\ layer and a \SI{3}{\nano\meter} GaAs cap were successively grown. We annealed the sample at \SI{850}{\celsius} for \SI{240}{\second} in N$_2$-atmosphere to improve the optical properties of the QDs.~\cite{moon12} A sketch of the cutaway profile of our sample is depicted in \figref{fig:sampleProfile}(a), while \figref{fig:sampleProfile}(b) shows a scanning electron microscope (SEM) image of a sample grown under identical conditions but uncapped, which revealed a QD density of 6--\SI{7}{\micro\meter^{-2}}. From atomic force microscopy (AFM) we found that the uncapped dots were lens-shaped and asymmetric in-plane with a major diameter of \SI{82.4\pm 7.6}{\nano\meter}, a minor diameter of \SI{54.4\pm 12.8}{\nano\meter} and a height of \SI{25.2\pm 8.8}{\nano\meter} but intermixing during overgrowth might change their size.~\cite{moon12}
In fact, we will show in sections~\ref{sec:OSQE} and \ref{sec:LOphonons} that ground-state excitons are strongly confined, which represents direct evidence of significant interdiffusion during annealing.

For optical measurements, the sample was placed in a liquid helium flow cryostat at \SI{10}{\kelvin} unless stated otherwise. A pulsed supercontinuum white-light source was spectrally filtered by an acousto-optic modulator at a wavelength of \SI{632}{\nano\meter} and was focused on the sample from the top to a spot size of about \SI{1.4}{\micro\meter^2} through a microscope objective with $\mathrm{NA}=0.6$. The wavelength corresponds to above-band excitation of the QDs. The emission from the QDs was collected by the same microscope objective. The cryostat was mounted on translation stages to control the excitation and collection spot with an accuracy of \SI{100}{\nano\meter}. The emission was spatially filtered by a circular aperture with a diameter of \SI{75}{\micro\meter} and was subsequently dispersed by a monochromator with a spectral resolution of \SI{50}{\pico\meter}. The filtered light was sent either to a charge-coupled device (CCD) for spectral measurements or to an avalanche photodiode (APD) for time-resolved measurements.

\section{Spectral Measurements}
\label{sec:OptEmission}

\begin{figure*}[t!]
	\includegraphics[width=\textwidth]{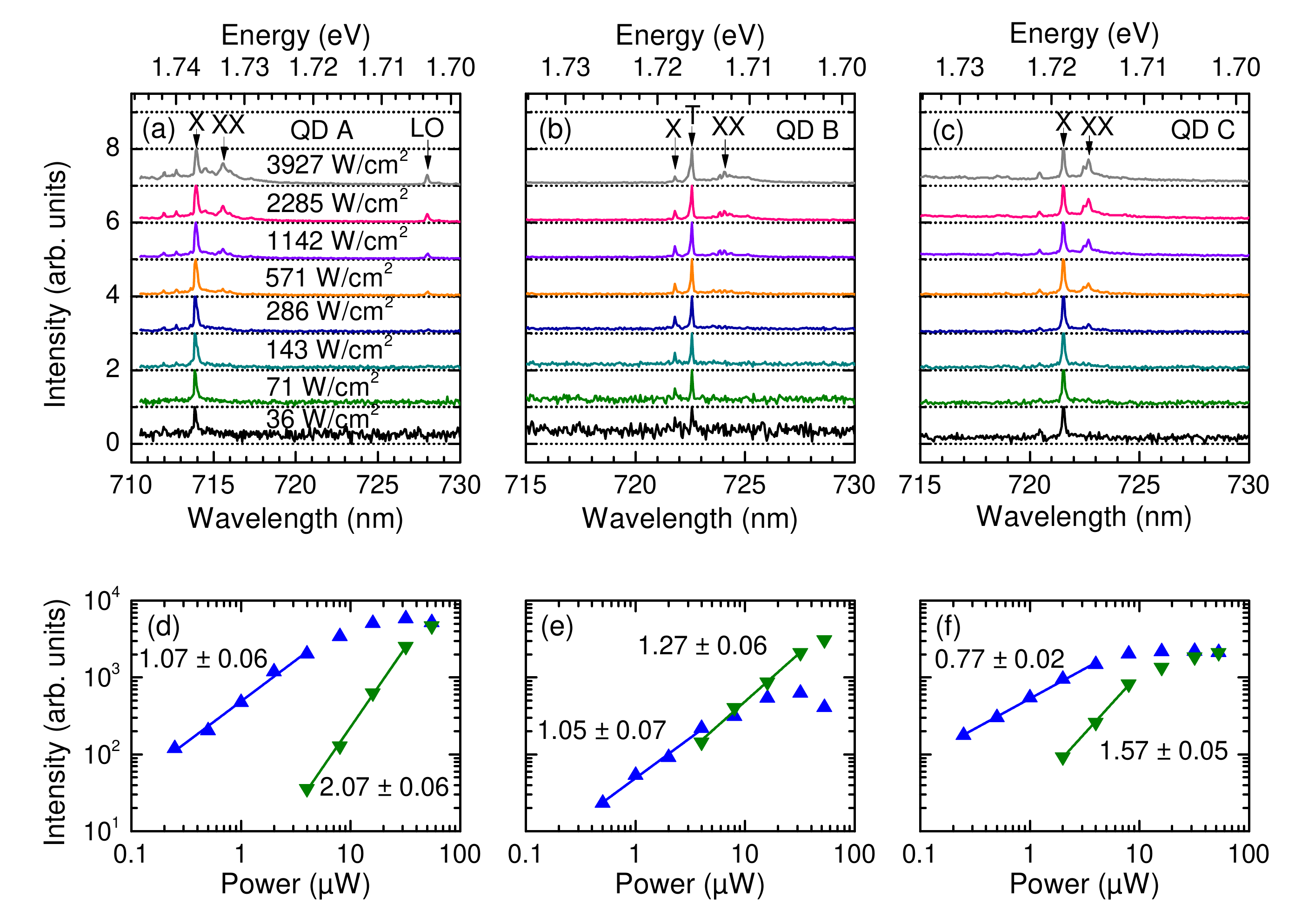}
	\caption{ \label{fig:optEmission} (Color online) (a-c)~Spectra at different excitation power densities for the three QDs discussed in this paper. The exciton, trion and biexciton lines are labelled as X, T and XX, respectively. For QD A, an LO-phonon replica is observed. (d-f)~Integrated intensity as a function of pumping power for the X (blue upward triangles) and XX lines (green downward triangles) along with the corresponding polynomial fits.  }
\end{figure*}

The optical properties of the QDs are investigated by means of above-band optical excitation, where electron-hole pairs are photoexcited in the \AlGaAs\ matrix in which the QDs are embedded. Due to the low areal density, the spectrum normally consists of individual lines at low average excitation power densities of \SI{143}{\watt\per\square{\centi\meter}} or below, which corresponds to the recombination of the ground-state exciton in the QD (further denoted as the X or exciton line, see \figref{fig:optEmission},(a) through (c)). For all three QDs, the integrated PL intensity of the X lines is approximately linear with excitation power, cf. Fig.~\ref{fig:optEmission}(d)-(f) as expected for excitons. The integrated intensity was calculated as follows. The X line was fitted with a Lorentzian plus a constant background whereupon the Lorentzian was integrated.
The exciton line saturates at a pumping intensity of about \SI{286}{\watt\per\square{\centi\meter}}, which corresponds to the onset of the biexciton as discussed later. The spectral behavior of QD B is different because two emission lines arise at low pumping powers. We have identified them as an exciton and a trion via time-resolved measurements, which is consistent with previous investigations of droplet-epitaxy GaAs QDs.\cite{abbarchi08_FS,kuroda02,kuroda09,abbarchi09} In particular, the exciton decays biexponentially (see \secref{sec:OSQE}) and the trion single-exponentially because it does not have a dark fine structure.\cite{bayer02}

The emission linewidth of QDs B and C is limited by the resolution of our spectrometer (\SI{50}{\pico\meter} equivalent to \SI{120}{\micro\electronvolt} at a wavelength of \SI{720}{\nano\meter}), which is clear evidence of single-QD emission. However, the line belonging to QD A is relatively broad and is found to be of the order of \SI{260}{\micro\electronvolt} after deconvolving it with the instrument response function, and is much broader than the radiative linewidth of several \si{\micro\electronvolt} of the excitonic transition in single QDs. This broadening is mainly caused by two factors. First, a noticeable broadening can be induced by the fine-structure splitting\cite{bester03} (FSS) of the bright exciton.\cite{abbarchi08} Second, the broadening of the exciton line is associated with spectral diffusion induced by a time-fluctuating quantum-confined Stark effect\cite{miller84} related to charging and discharging of trap defects.\cite{berthelot06} This scenario is plausible, given the low-temperature growth of the droplet-epitaxy QDs, which can affect the quality of their crystalline structure.

At higher excitation powers (\SI{286}{\watt\per\square{\centi\meter}} or above) we observe a second line, which is redshifted by 2--\SI{4}{meV} with respect to the exciton line. From power-series measurements (see \figref{fig:optEmission}), the slope of the integrated intensity (the raw data were integrated directly due to the broad and irregular shape of the line) is significantly larger than that of the X-line, which suggests a biexciton to exciton recombination. For biexcitons, the PL intensity is expected to be quadratic with excitation power. For QD A the data show good agreement but for QDs B and C the slope of the second peak is found to be superlinear but less than two. Therefore, we performed lifetime measurements on these two peaks to confirm the biexcitonic origin of the second peak. The details about decay curve modelling will be extensively discussed in the following section; for now, we will just use the results. We obtain a total decay rate for the exciton (secondary) line lying in the range 1.6--\SI{2.1}{\nano\second^{-1}} (3.6--\SI{4.3}{\nano\second^{-1}}) for all three QDs. In the limit of slow spin-flip processes,\cite{narvaez06} the biexciton should be twice as fast as the exciton because it has two possible radiative decay channels, i.e., it can decay to either of the bright states, while the exciton has a single radiative decay channel. Similarly, the biexciton is expected to decay twice as fast nonradiatively because any of its charge carriers is prone to nonradiative loss. For example, let us assume that one type of charge carriers (e.g., holes) have the largest nonradiative decay, and that the hole decays with the same rate from an exciton or biexciton. Then, in the first approximation, an exciton $\ket{\uparrow \Downarrow}$ decays to $\ket{\uparrow 0}$, while a biexciton $\ket{\uparrow \downarrow \Uparrow \Downarrow}$ can decay either to $\ket{\uparrow \downarrow \Uparrow 0}$ or to $\ket{\uparrow \downarrow 0 \Downarrow}$, where the single arrow denotes the electron spin and the double arrow the hole spin. As a consequence, the total decay rate of the biexciton is expected to be $\gamma_{XX} = 2\gamma_{X,\mathrm{RAD}} + 2\gamma_{X,\mathrm{NRAD}} = 2\gamma_{X}$, where $\gamma_{X}$ is the total decay rate of the exciton. The time-resolved measurements confirm that the second line is due to biexciton recombination (further denoted as the XX line).

The XX saturation intensity should correspond to the onset of multi-particle recombination because more than four single-particles (2 electrons + 2 holes) are stored in the QDs. Indeed, an increasing number of spectral lines on top of a continuous background appear at yet larger excitation intensities. Additionally, the XX line exhibits a peculiar feature: it is spectrally broadened and time-resolved measurements show that the low-energy sideband decays first. In fact, this behavior was seen in the past for GaAs interface-fluctuation QDs at large excitation powers and was attributed to Coulomb interaction between the biexciton and charge carriers present at higher lying states in the surrounding quantum well.\cite{senellart05} Our scenario is in fact very similar, the main difference being that the higher lying states belong to the same QD and not to a quantum well.

Aside from these common features, each QD has its own spectral repertoire. An interesting example is the line of QD A at a wavelength of \SI{728}{nm}. Its energy distance to the exciton peak is \SI{33.4}{\milli\electronvolt}, which suggests an optical phonon replica since the bulk GaAs LO (TO) phonon energy is \SI{36.6}{\milli\electronvolt} (\SI{33.2}{\milli\electronvolt}). By fitting the emission spectrum with a Lorentzian function and deconvolving with the point-spread function of the setup we obtain a FWHM of \SI{430}{\micro\electronvolt}. At this particular excitation power (far beyond saturation), the X line is \SI{436}{\micro\electronvolt} broad. Consequently, we attribute this red-shifted spectral emission to the optical phonon replica of the exciton.

\begin{figure}[t!]
	\includegraphics[width=0.4\textwidth]{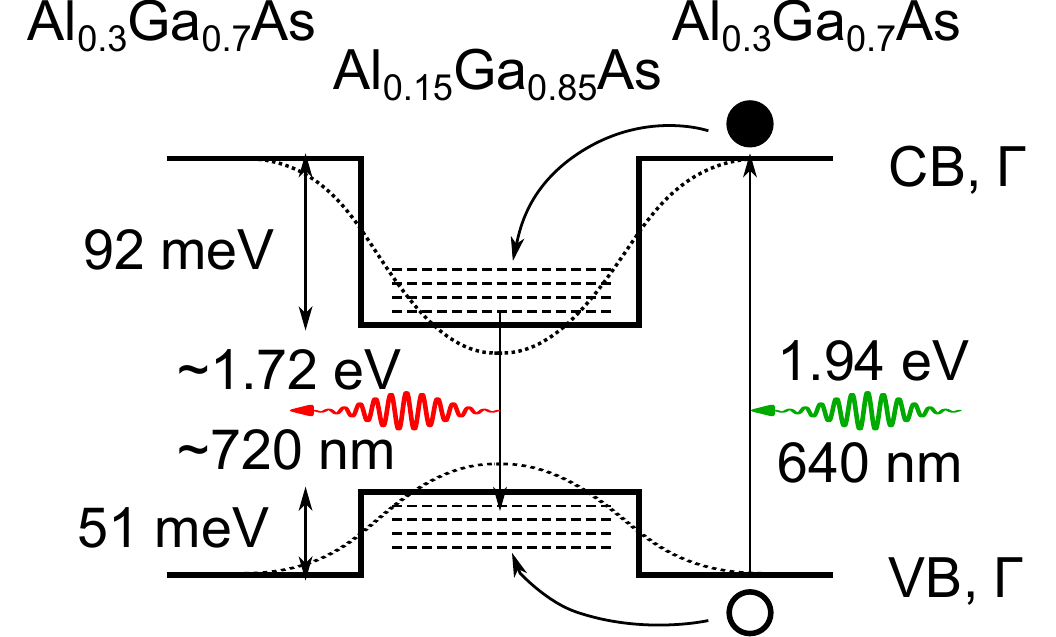}
	\caption{ \label{bandStruct} (Color online) Approximated band diagram (solid black lines) of the QDs under the assumption of constant Al content within the QD. Charge carriers are generated inside the \AlGaAs\ matrix and are subsequently trapped by the Al$_{0.15}$Ga$_{0.85}$As QDs before recombining radiatively around a wavelength of \SI{720}{\nano\meter}. The dotted line is a qualitative sketch of the actual potential profile whose smooth spatial dependence is a consequence of alloy inhomogeneities within the QD, thereby rendering the spatial extent of the ground-state exciton smaller than the size of the QD (see text for details).}
\end{figure}

To conclude this section, let us calculate the band structure of the QDs. The only information taken from experiment is the emission frequency of the exciton. The QDs emit in a wavelength range between 700 and \SI{740}{\nano\meter} and the band gap of GaAs (\AlGaAs) of about \SI{820}{\nano\meter} (\SI{640}{\nano\meter}) at low temperatures,\cite{vurgaftman01} and given the fact that confinement effects are supposed to be small due to the large size of the QDs, we conclude that the growth has resulted in a substantial interdiffusion between the AlGaAs matrix and the GaAs QDs. Consequently, we expect the conduction and valence band potential profiles to follow the intermixing profile, as is qualitatively sketched in \figref{bandStruct}. Unfortunately we do not know the explicit spatial dependence of the latter; therefore, in order to provide a quantitative picture of the average interdiffusion magnitude, we assume for the moment that the potential profile is constant. This is of course a drastic assumption and is just meant to provide a simplified picture of the band structure and, therefore, its consequences should be treated with care (in fact, we will see later that the potential profile does exhibit a spatial dependence). Similarly, by virtue of the previous arguments, we believe that confinement effects are smaller than the involved energy scales of the band diagram, therefore we neglect them to simplify the discussion. Then, in the effective-mass approximation, we can write down the energy position in eV of the conduction $E_{c,\Gamma}$ and valence $E_{v,\Gamma}$ bands of Al$_x$Ga$_{1-x}$As at the $\Gamma$ point in Fourier space\cite{vurgaftman01,zunger98}
\begin{align}
E_{c,\Gamma}(x) &= 2.979 + 0.765x + 0.305 x^2,\\
E_{v,\Gamma}(x) &= 1.460 - 0.509x.
\label{eq:indivBands}
\end{align}

By solving for $E_{c,\Gamma}-E_{v,\Gamma}=E_\mathrm{PL}$, where $E_\mathrm{PL}$ is the emission energy, we obtain an average Al content of \SI{15.3}{\percent} for a wavelength of \SI{720}{\nano\meter}. This yields a total confinement energy of \SI{51}{\milli\electronvolt} for holes and \SI{92}{\milli\electronvolt} for electrons. The corresponding band diagram, which includes the aforementioned simplifications, is sketched in \figref{bandStruct}. We would like to underline that the first valence band eigenstate is expected to be heavy-hole like due to the relatively small aspect ratio of the QDs\cite{huo12} as seen in \secref{sec:ExpProc} (diffusion should not change the aspect ratio because it is approximately isotropic). This can be understood by taking the limit of vanishing in-plane confinement. Then, the quantization axis would be the same (along the $z$-axis, parallel to the QD height)\cite{coldren12} and the heavy-hole band would completely decouple from split-off and light-holes at the $\Gamma$-point and would be energetically closest to the conduction band, as is well-known for quantum wells.\cite{harrison05} The presence of a finite in-plane confinement, as is the case for small aspect-ratio QDs, induces a small contribution from the light-hole band, which can be neglected for practical purposes.\cite{huo12}

\section{Oscillator Strength and Quantum Efficiency}
\label{sec:OSQE}

Spectral measurements provide important insight to the level structure of QDs, as we have seen in the previous section. However, phenomena with a lifetime significantly shorter than a few hundred milliseconds are averaged out and therefore not resolved. For instance, spontaneous emission, spin-flip processes, phonon scattering, etc., are processes which occur somewhere between picosecond to microsecond time scales. Time-resolved measurements of the PL signal have the capability of providing rich information about such processes. The figure of merit quantifying the coupling of an emitter to light is the \OS\ $f$, which is proportional to the radiative decay rate in a homogeneous medium $\gamma_\mathrm{RAD}^\mathrm{HOM}$ via\cite{stobbe12}
\begin{equation}
f = \frac{6 \pi m_0 \epsilon_0 c_0^3}{e^2n\omega_0^2}\gamma_\mathrm{RAD}^\mathrm{HOM},
\label{eq:OSdef}
\end{equation}
where $n$ is the refractive index of the host material (in our case \AlGaAs ), $\omega_0$ and $c_0$ constitute the frequency and speed of light, respectively, $\epsilon_0$ the vacuum permittivity, $m_0$ the electron mass, and $e$ the elementary charge. In other words, the oscillator strength of the QD can be obtained by measuring the radiative decay rate of the ground-state exciton. However, the latter is not a straightforward task because the measured decay rate $\gamma_\mathrm{TOT}$ is the sum of the radiative rate $\gamma_\mathrm{RAD}$ and all nonradiative recombination channels $\gamma_\mathrm{NRAD}$. In general, nonradiative processes are omnipresent in solid-state systems even at low temperatures and, therefore, cannot be neglected. We can define the intrinsic quantum efficiency $\eta$ as the probability of an emitter to recombine radiatively
\begin{equation}
\eta = \frac{\gamma_\mathrm{RAD}^\mathrm{HOM}}{\gamma_\mathrm{RAD}^\mathrm{HOM}+\gamma_\mathrm{NRAD}}.
\label{eq:QEdef}
\end{equation}
In this section we extract the \OS\ and \QE\ of droplet-epitaxy QDs by accounting for the transfer of the excitonic population within the QD.

\begin{figure*}[t!]
    \includegraphics[width=0.48\textwidth]{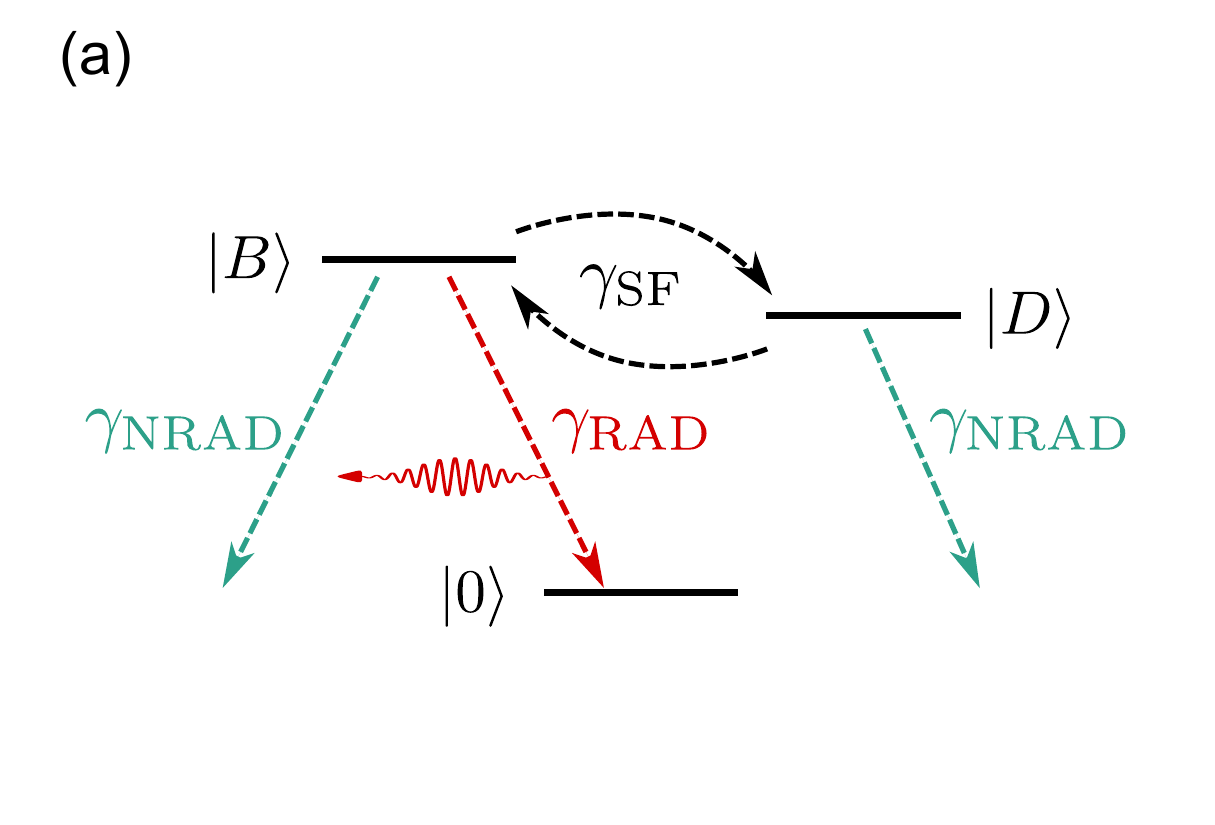}~
  	\includegraphics[width=0.48\textwidth]{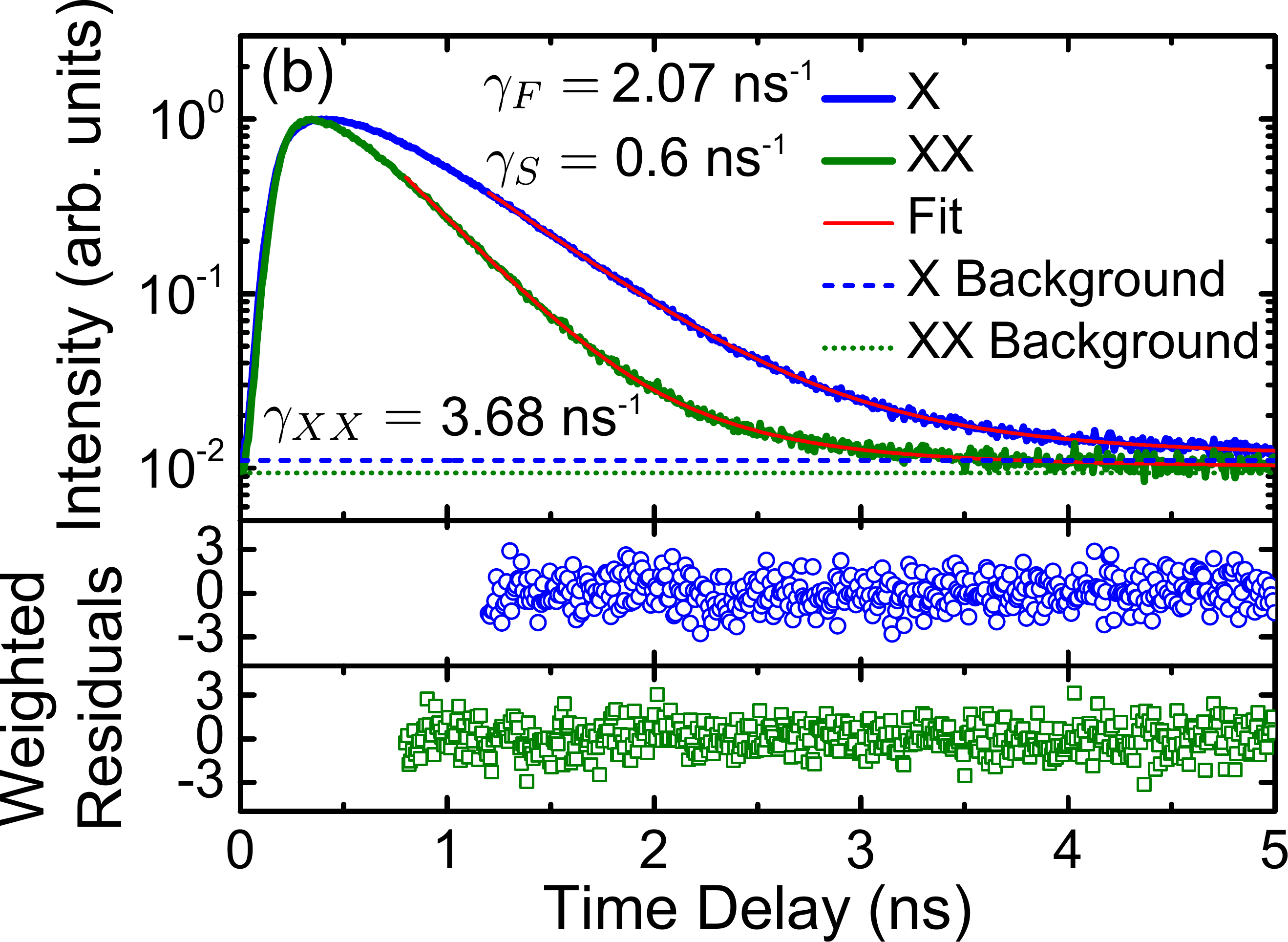}
    \caption{ \label{fig:decayDynamics} (Color online) (a)~Level scheme describing the relevant population transfer of the exciton in a QD, which has a direct impact on the bright exciton decay dynamics. The bright exciton $\ket{B}$ can decay either radiatively ($\gamma_\mathrm{RAD}$) or nonradiatively ($\gamma_\mathrm{NRAD}$) and can interact with its dark counterpart $\ket{D}$ via the spin-flip rate ($\gamma_\mathrm{SF}$). (b)~Time-resolved decay dynamics of the exciton (blue) and biexciton (green) of QD A along with the corresponding weighted residuals marked by blue circles and green squares, respectively. The red solid line indicates the fit while the dashed/dotted lines denote the background level. The data are fitted $\sim$\SI{0.5}{\nano\second} later than the beginning of the decay due to filling effects (see text).}
\end{figure*}

If the excitation intensity is below the onset of the biexciton line (see \figref{fig:optEmission}), only one electron and one hole are captured by the QD. These carriers undergo phonon-scattering processes on a time scale of the order of picoseconds (several orders of magnitude faster than the radiative recombination) before they end up in the QD ground state. We employ the effective-mass approximation to describe the QDs.\cite{miller08} We have discussed in \secref{sec:OptEmission} that the hole ground state is to a large extent heavy-hole like with a total angular momentum $J_h=3/2$ and two possible projections $J_{h,z}=\pm 3/2$,\cite{cardona10} which we refer to as $\ket{3/2,\pm 3/2}$. 
The conduction band Bloch functions are spherically symmetric and, therefore, the electron has no orbital angular momentum. Hence, its total angular momentum is given by its spin with $\ket{1/2,\pm 1/2}$. Consequently, four exciton eigenstates are formed with $J=2$ and $J_z=\left\{ \pm 1, \pm 2 \right\}$. Two excitons out of four have a projected angular momentum of $\pm 1$ and are optically bright, which is why they are referred to as bright excitons $\ket{B}$, see \figref{fig:decayDynamics}(a). Excitons with $J_z=\pm 2$ do not couple to the light field and are denoted as dark excitons $\ket{D}$. Due to electron-hole exchange interaction, bright excitons are situated higher in energy than dark excitons by an amount $\Delta_\mathrm{BD}$ of several hundred \si{\micro\electronvolt}. \cite{bester03}

The exciton captured by the QD ends up either being bright or dark with equal likelihood because above-band excitation is performed.\cite{baylac95} Dark excitons $\ket{D}$ cannot decay radiatively but they can decay nonradiatively with the corresponding nonradiative recombination rate $\gamma_\mathrm{NRAD}$, as seen in \figref{fig:decayDynamics}(a). Dark excitons can become bright through spin-flip processes at the rate $\gamma_\mathrm{SF}$. Unlike for bulk semiconductors or quantum wells, which have quasi-continuous energy states, the energy levels in QDs are quantized, which makes it difficult for charge carriers to simultaneously flip spin and fulfil energy conservation.\cite{paillard01,woods02} This leads to two important consequences. First, spin-flip processes are primarily phonon-mediated owing to the energy conservation requirements. This is possible because the thermal energy $k_B T \gg \Delta_\mathrm{BD}$ at the temperatures the experiment is carried out, and this is the reason why dark-bright and bright-dark flip rates are assumed to be the same. And second, QDs are characterized by slow spin-flip rates and have been measured to be of the order of hundred nanoseconds, i.e., much slower than radiative rates.\cite{johansen10} It is important to underline that spin flip is the key process allowing us to separate radiative from nonradiative rates, as we will see shortly. In turn, the bright state $\ket{B}$ can decay both radiatively with $\gamma_\mathrm{RAD}$ and nonradiatively with $\gamma_\mathrm{NRAD}$, and it can also become dark by a spin-flip process $\gamma_\mathrm{SF}$. The nonradiative rates are taken to be the same for bright and dark excitons due to their small energy splitting.\cite{johansen10} Consequently, we can write the equations governing the transfer of the excitonic population within the QD for pulsed excitation below saturation
\begin{equation}
\label{eqn:systemEqs}
\begin{pmatrix}
\dot{\rho}_B\\
\dot{\rho}_D
\end{pmatrix}
=
\begin{pmatrix} 
-\gamma_\mathrm{RAD}-\gamma_\mathrm{NRAD}-\gamma_\mathrm{SF} & \gamma_\mathrm{SF}\\
\gamma_\mathrm{SF} & -\gamma_\mathrm{NRAD}-\gamma_\mathrm{SF}
\end{pmatrix}
\begin{pmatrix}
\rho_B\\
\rho_D
\end{pmatrix},
\end{equation}
where $\rho$ denotes the probability to occupy the corresponding level and the dot indicates the time derivative. Under the realistic assumption that spin flip-processes are much slower than the radiative decay rate, i.e., $\gamma_\mathrm{SF} \ll \gamma_\mathrm{RAD}$, we solve \eqref{eqn:systemEqs} and obtain the temporal decay of the bright state
\begin{equation}
\rho_B(t) = \rho_B(0)e^{-(\gamma_\mathrm{RAD}+\gamma_\mathrm{NRAD})t} + \frac{\gamma_\mathrm{SF}}{\gamma_\mathrm{RAD}}\rho_D(0)e^{-(\gamma_\mathrm{NRAD}+\gamma_\mathrm{SF})t}.
\end{equation}
The bright exciton exhibits a biexponential decay with the fast rate $\gamma_F = \gamma_\mathrm{RAD} + \gamma_\mathrm{NRAD}$ and the slow rate $\gamma_S = \gamma_\mathrm{NRAD} + \gamma_\mathrm{SF}$. Consequently, by fitting the measured decay curves with $f(\tau) = A_F e^{-\gamma_F\tau} + A_S e^{-\gamma_S\tau} + C$, where $\tau$ is the time delay with respect to the start of the excitation pulse and $C$ is the background level, which is determined by the measured dark-count rate and after-pulsing probability of the detector, we can unambiguously extract the radiative and nonradiative rates via
\begin{align}
\gamma_\mathrm{RAD}  &= \gamma_F - \gamma_S + \frac{A_S}{A_F}\frac{\rho_B(0)}{\rho_D(0)},\\
\gamma_\mathrm{NRAD} &= \gamma_S - \frac{A_S}{A_F}\frac{\rho_B(0)}{\rho_D(0)}.
\label{eq:extractRAD}
\end{align}
Due to the fact that above-band excitation is performed, $\rho_B(0)/\rho_D(0)=1$ since uncorrelated electron-hole pairs are generated that are equally likely to be prepared in either of the excitonic states. The extracted radiative rate $\gamma_\mathrm{RAD}$ does generally not equal the homogeneous radiative rate $\gamma_\mathrm{RAD}^\mathrm{HOM}$ because the emitter is not placed in an infinite homogeneous medium. Therefore, we calculate the normalized LDOS at the position of the emitter~\cite{paulus00} for the layered structure outlined in \figref{fig:sampleProfile}(a) and obtain a value of 1.05. We use this value along with Eqs.~(\ref{eq:OSdef}--\ref{eq:QEdef}) to extract the \OS\ and \QE\ of droplet-epitaxy QDs.

Experimentally, the decay dynamics is recorded by selecting a single spectral line and sending it to the APD for time-resolved measurements. The excitonic and biexcitonic decays of QD~A, both below saturation, are plotted in \figref{fig:decayDynamics}(b) along with the corresponding fits. In order to quantify how well the fit reproduces the experimental data we define the weighted residual $W_k$ as
\begin{equation}
W_k = \frac{\rho_M(t_k)-\rho_F(t_k)}{\sqrt{\rho_M(t_k)}},
\end{equation}
where $\rho_M$ is the measured data, $\rho_F$ represents the fitted value, and the discreetness of the time-delay axis is denoted by the subscript $k$. The biexponential decay is fitted to the acquired data using a least-squares approach where the collapsed residual $\chi_R^2 = \frac{1}{N-p}\sum_{k=1}^{N} W_k^2$ is minimized, $N$ being the total number of time bins and $p$ the number of adjustable parameters in the model.

The exciton exhibits a biexponential behavior, which is confirmed by the low $\chi_R^2$ (see Table~\ref{table:decayParameters}; it is important to emphasize that a single exponent severely underfits all the decay curves at \SI{10}{\kelvin}) with the fast rate of the order of \SI{2}{\nano\second^{-1}} and the slow rate about three times smaller, as can be seen in \figref{fig:decayDynamics}(b). We extract an \OS\ of around 9 and a \QE\ between 69 to \SI{79}{\percent} (see Table~\ref{table:decayParameters}). Even though the \QE\ of droplet-epitaxy QDs is found to be lower than that of small InAs QDs, their optical quality is significantly higher than that of large InAs QDs whose \QE\ ranges between 30 and \SI{60}{\percent}.\cite{stobbe10} We believe the reason for this to be the absence of strain-related effects in GaAs QDs, which makes \DE\ a growth technique potentially capable of delivering QDs with very high optical quality suitable for quantum-information applications. We attribute the less-than-unity \QE\ to the low-temperature growth of the capping layer, see \secref{sec:ExpProc}. 

\begin{table}[t!]
\begin{center}
\begin{tabular}{l c c c}\botrule
                                                   & QD A         & QD B     & QD C  \\ \colrule
$\gamma_F$ (\si{\nano\second^{-1}})				   & 2.07         & 1.61     & 2.00  \\
$\gamma_S$ (\si{\nano\second^{-1}})                & 0.60         & 0.34     & 0.62  \\
$A_S/A_F\times10^{-3}$                             & 10           & 50       & 23    \\
$\chi_R^2$                                         & 1.1          & 1.05     & 1.02  \\ \colrule
$\gamma_\mathrm{RAD}$ (\si{\nano\second^{-1}})     & 1.47         & 1.27     & 1.38  \\
$\gamma_\mathrm{NRAD}$ (\si{\nano\second^{-1}})    & 0.58         & 0.28     & 0.59  \\
$\gamma_\mathrm{SF}$ (\si{\micro\second^{-1}})     & 14.4         & 57.7     & 30.4  \\
\OS                                    			   & 9.4          & 8.2      & 9.0   \\
\QE (\%)                                           & 70.1         & 78.1     & 69.0  \\
$\abs{\left< \psi_h | \psi_e \right>}^2$ (\%)      & 56.5         & 49.1     & 53.5  \\
\botrule
\end{tabular}
\end{center}
\caption{Quantities extracted from the exciton decay.}
\label{table:decayParameters}
\end{table}

It is commonly stated that a key advantage of QDs relies in their \OS, which is about one order of magnitude larger than that of atomic emitters. However, it is important to underline that the \OS\ depends on the QD size. Only for QDs smaller than the exciton Bohr radius $a_0$ (further denoted as `small QDs') does the \OS\ become almost independent of the QD size.~\cite{hanamura88,stobbe12,Einevoll1992} In the dipole- and effective-mass approximations, the \OS\ of small QDs is given by 
\begin{equation}
f = \frac{E_P}{\hbar \omega_0} \abs{\bra{\psi_h}\left. \psi_e \right>}^2
\label{eq:OSstrongconf}
\end{equation}
where $E_P$ is the Kane energy and $\psi_e$ ($\psi_h$) is the electron (hole) slowly-varying envelope function. This so-called `strong confinement regime' has an upper bound for the \OS\ of $f_\mathrm{MAX}=E_P/\hbar\omega_0$, which amounts to 16.7 for a GaAs QD at an emission wavelength of \SI{720}{\nano\meter}, where we have used a GaAs value of \SI{28.8}{\electronvolt} for the Kane energy.\cite{vurgaftman01}

On the other hand, QDs whose linear size $L$ is larger than $a_0$ (further denoted as `large QDs') exhibit an enhanced light-matter interaction. For example, the oscillator strength of a spherical QD is given by\cite{stobbe12} $f_\mathrm{SPH} = f_\mathrm{MAX} \times \sqrt{\pi}\left( L/a_0 \right)^3$, and scales with the number of unit cells the exciton spreads itself across. The \OS\ in weakly-confined systems can become significantly larger than $f_\mathrm{MAX}$ if $L>a_0$. This behavior of large QDs was coined the giant oscillator strength, and its physical reason is related to the superradiant nature of the ground-state exciton, which distributes itself coherently over a much larger volume than it otherwise does in small QDs or bulk.~\cite{hanamura88}

According to the AFM data, the QDs have an in-plane radius of 30--\SI{40}{\nano\meter} and a height of \SI{25}{nm} before capping and annealing. Given the fact that all the dimensions are weakly confined, we compare the QD to a sphere with the same volume and obtain an expected \OS\ beyond 900, which is two orders of magnitude larger than the observed oscillator strengths of about 10 that are listed in Table~\ref{table:decayParameters}. This value is within the strong confinement limit, which is direct evidence that the excitons are strongly confined in the droplet-epitaxy QDs. In other words, it appears that the effective size of the QDs is diminished by the capping and annealing processes. This assumption is supported by the emission wavelength (see the discussion in \secref{sec:OptEmission}), which is substantially smaller than the GaAs bandgap, thereby suggesting considerable alloy inhomogeneities in the QDs and, hence, a reduction of the ground-state exciton coherence volume. Our results clearly underline the importance of a growth technique which would induce as little intermixing as possible between the QD and the surrounding matrix in order to obtain enhanced light-matter interaction. With the help of \eqref{eq:OSstrongconf} we calculate the electron-hole overlap integral $\abs{\bra{\psi_h}\left. \psi_e \right>}^2$ to range between 0.49--0.57, which is comparable but smaller than that of SK InAs QDs (0.62--0.77).\cite{johansen08}

Let us return to the decay dynamics of the biexciton, as shown in \figref{fig:decayDynamics}(b). Surprisingly, a single-exponential does not fit the curve for all the investigated QDs. We attribute this to spectral pollution of the biexciton line by LA phonons stemming from the exciton line.\cite{besombes01,favero03,peter04} Hence, the PL signal $I_\mathrm{APD}$ at the emission frequency of the biexciton within a frequency range determined by the resolution of the setup (\SI{50}{\pico\meter}) is given by
\begin{equation}
I_\mathrm{APD} = A_{XX}e^{-\gamma_{XX}t} + C_{LA}A_Fe^{-\gamma_Ft} + C_{LA}A_Se^{-\gamma_St} + C,
\label{eq:XXdecay}
\end{equation}
where $C_{LA}$ is the integrated coupling coefficient of the zero-phonon line to the modes which overlap spectrally with the biexciton emission. It is clear from \eqref{eq:XXdecay} that the decay curve is expected to be triple-exponential with the fastest rate $\gamma_{XX}$ corresponding to the biexciton decay rate.

We have fitted the biexciton curve with a triple exponential, see the red solid line in \figref{fig:decayDynamics}(b). The extracted fast rate $\gamma_{XX}=\SI{3.68}{\nano\second^{-1}}$ roughly equals $2\gamma_\mathrm{F}$ of the bright exciton (see Table~\ref{table:decayParameters}), in good agreement with the theoretical considerations in \secref{sec:OptEmission}. The middle and slow rates are found to be \SI{2.2}{\nano\second^{-1}} and \SI{0.44}{\nano\second^{-1}}, respectively, and reproduce quite accurately the fast and slow rates of the exciton line (noteworthy, the biexciton line was recorded above the exciton saturation, which is why the decay rates of the exciton at this elevated power might be different than the ones given in Table~\ref{table:decayParameters}), which brings further evidence of a phonon-mediated emission of the ground-state exciton overlapping spectrally with the biexciton.

To conclude this section, we emphasize that although the ground state is clearly strongly confined, the excited states do not necessarily have to be so. We have already shown that it is very likely that the droplet-epitaxy QDs are characterized by a smooth spatial potential, which follows the alloy-intermixing profile implying that excited states become less confined (see \figref{bandStruct}). In general, there is no obvious correlation between the confinement of ground-state excitons and the QD size, if the potential profile is not uniform within the QD. In fact, despite the strong confinement of the ground state, the droplet-epitaxy QDs can be considered `large' because they contain a large number of excited states, which is supported by two independent experimental findings. First, time-resolved measurements of the exciton line above saturation show pronounced filling effects (i.e., there is a substantial time interval between the excitation pulse and the actual PL decay), which is characteristic for large QDs.\cite{senellart05,stobbe10} Second, the effective transition strength is diminished with increasing temperature, which is direct evidence of nearby excited states (several meV away), as is shown in the following section.

\section{Temperature Dependence of the Effective Transition Strength}
\label{sec:TOS}

\begin{figure*}[t!]
	\includegraphics[width=\textwidth]{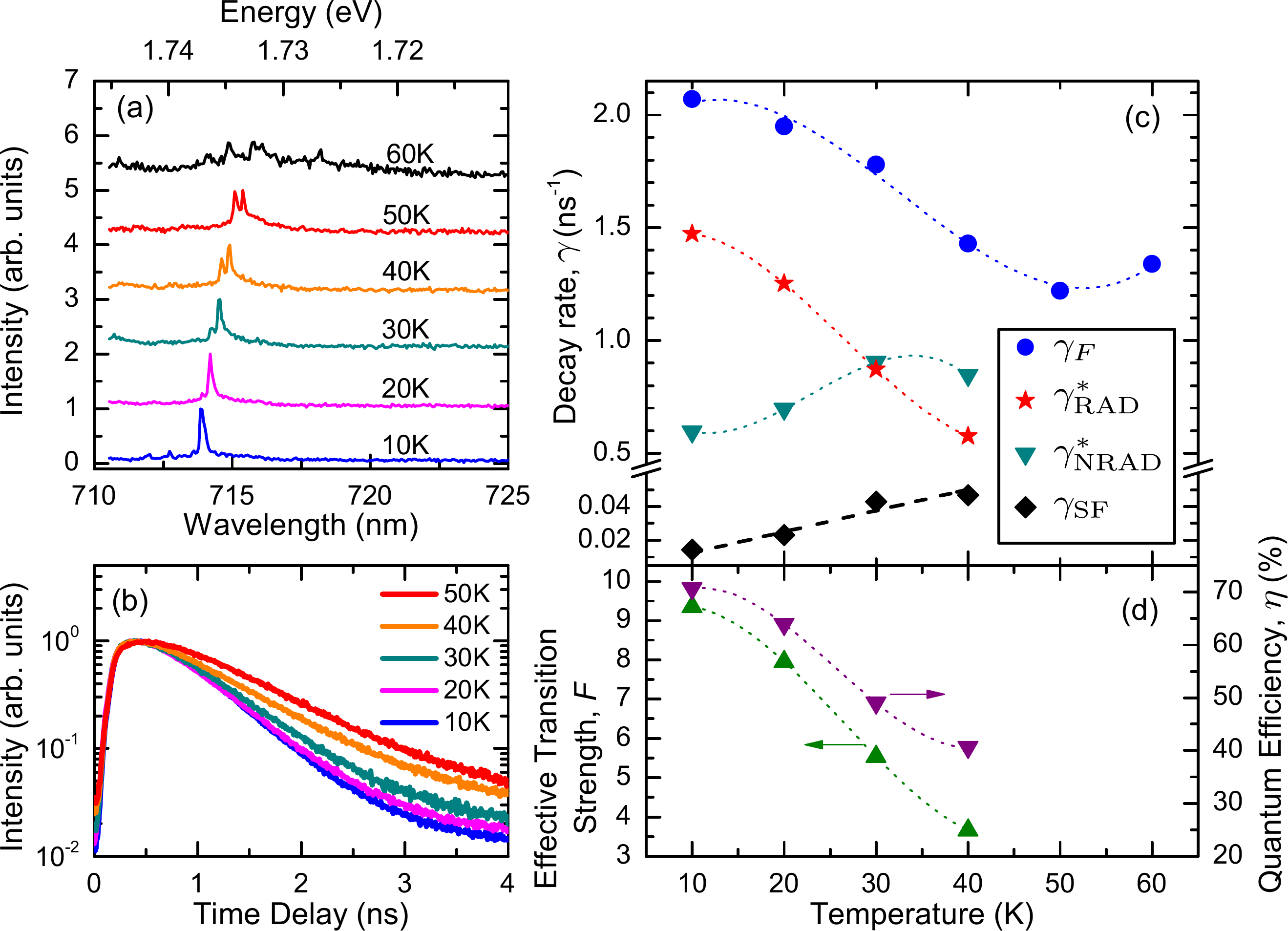}
    \caption{ \label{fig:dotA_temp} (Color online) Spectra and decay dynamics of excitons captured by QD A at various temperatures for an excitation power density of \SI{286}{\watt\per\square\centi\meter}, which is below saturation of the exciton line. (a)~Spectra recorded within a temperature range of 10--\SI{60}{\kelvin}. (b)~Time-resolved decay of the exciton line from \SI{10}{\kelvin} to \SI{50}{\kelvin} showing an increase in the exciton lifetime with temperature. (c)~Temperature dependence of the fast, radiative, nonradiative, and spin-flip rates of the exciton, as well as (d) its effective transition strength and \QE. The black dashed line fits the spin-flip rates with a linear function passing through the origin. The dotted lines provide guides to the eye.}
\end{figure*}

Due to three-dimensional confinement, QDs have discrete energy levels. At low excitation powers and temperatures, only the ground state is relevant because the first excited eigenstate is situated at much higher energies than the thermal energy $k_B T$. This picture is justified for small QDs where confinement effects are significant but it might no longer be valid in large QDs where the energy difference between the eigenstates may become comparable to the thermal energy; for example, $k_BT=\SI{4.3}{\milli\electronvolt}$ at \SI{50}{\kelvin}. Thermal population of excited states leads to a modification of the level scheme in \figref{fig:decayDynamics}(a) and, thus, of the exciton dynamics. More importantly, if a single excitation is thermally shared by several eigenstates, the effective transition strength becomes temperature-dependent and does not coincide with the \OS\ of the ground-state exciton.~\cite{feldmann87} Generally, the oscillator strength is a property of two energy levels and quantifies the emission rate of light. When an exciton is shared among many energy levels (as in the case of a quantum well at finite temperatures for example), the radiative decay rate of the system can no longer be used to extract the oscillator strength. In this context, the effective transition strength becomes a more relevant quantity and determines the light emission rate.\cite{feldmann87} In the following, we present a study of the temperature properties of droplet-epitaxy QDs.

Some of the QDs we have investigated exhibit a striking reduction in their effective transition strength as the temperature is increased. In this section we show that this behavior is a direct consequence of their large size. On the other hand, some QDs do not have a pronounced temperature dependence of the decay dynamics, which suggests a limited interaction with excited states or, alternatively, a smaller effective size. The latter are not considered further in the following.

Figure~\ref{fig:dotA_temp}(a) displays the acquired PL spectrum of QD A in a temperature range from 10 to \SI{60}{\kelvin} below the exciton saturation. A pronounced redshift of the excitonic line is observed due to the well-known band gap shrinkage with temperature. At \SI{60}{\kelvin} the PL signal is quenched, probably due to the low quantum efficiency of the transition. A narrow line, blueshifted by \SI{700}{\micro\electronvolt} with respect to the exciton line, appears with increasing temperature. It cannot be an excited state because it would have been thermally populated at energies $4k_BT \approx \SI{700}{\micro\electronvolt}$ corresponding to $T \approx \SI{2}{\kelvin}$. Time-resolved measurements revealed that it decays identically to the exciton line for all temperatures. This is consistent with the behavior of a charged exciton (a trion), which is expected to decay with roughly the same rate as the neutral exciton. Henceforth we turn our attention exclusively to the exciton line.

QD A reveals a pronounced dependence of the decay dynamics on temperature, see \figref{fig:dotA_temp}(b). As the temperature is increased, the bright exciton decays slower up to \SI{50}{\kelvin}. Interestingly, the decay curves become single-exponential at temperatures higher than \SI{40}{\kelvin}. This is a consequence of $\gamma_\mathrm{NRAD}$ becoming comparable to $\gamma_\mathrm{RAD}$, whereby the biexponential decay is masked by the measurement noise and the curves become single-exponential. More formally: in order to resolve a biexponential decay, the PL signal of the fast component must decay before the amplitude of the slow component becomes smaller than the background noise $\Delta A_\mathrm{BG}$, i.e., $A_Fe^{-\gamma_Ft} \leq A_se^{-\gamma_St}$ and $\Delta A_\mathrm{BG} < A_se^{-\gamma_St}$. This gives the important condition for experimentally observing the slow decay component
\begin{equation}
\frac{\gamma_\mathrm{NRAD}}{\gamma_\mathrm{RAD}} \lesssim \frac{\ln\left(\frac{A_S}{\Delta A_\mathrm{BG}}\right)}{\ln\left(\frac{A_F}{A_S}\right)}.
\label{eq:seeSlow}
\end{equation}
In the limit of noiseless measurements, the slow component can always be detected, whereas if the noise equals the slow component amplitude, the biexponential decay cannot be resolved at all and the curve appears single-exponential. It is also clear that a longer integration time $\tau$ of the decay curves enables resolving the biexponential decay better because the PL signal scales with $\tau$ and the measurement noise with $\sqrt{\tau}$. In our experiment, $A_S/\Delta A_\mathrm{BG}~\approx~60$ and $A_F/A_S~\approx~20$ at a temperature of \SI{40}{\kelvin} for example, which yield $\gamma_\mathrm{NRAD}/\gamma_\mathrm{RAD}~\lesssim~1.4$. Indeed, at \SI{40}{\kelvin} $\gamma_\mathrm{NRAD}/\gamma_\mathrm{RAD}$ is about 1.4, cf. \figref{fig:dotA_temp}(c), while at \SI{50}{\kelvin} the decay cannot be fitted with a biexponential, which is direct evidence of the low quantum efficiency of the transition. This is confirmed by the luminescence quenching in the spectrum, see \figref{fig:dotA_temp}(a). Henceforth only the biexponential curves will be discussed.

The sudden threefold drop in the radiative decay rate and, consequently, in the effective transition strength with temperature (see \figref{fig:dotA_temp}(c) and (d)) is puzzling at first sight since the spontaneous emission rate of QDs is expected to be independent of temperature at low temperatures.\cite{fiore00} It is worth noting that a quantum well for example does decay slower with increasing temperature owing to thermal excitation of excitons away from the Brillouin zone center, which makes them optically dark.\cite{butov99,koch06} A similar effect was predicted theoretically for QDs and attributed to thermal population of excited states\cite{citrin93} but is not expected to occur in small QDs at low temperatures by virtue of their zero-dimensional density of states. In large QDs, however, thermal excitation may occur at substantially lower temperatures. We will elaborate on this in the next paragraphs. We would like to point out that a reduction in the decay rate was observed for SK InAs QDs in a similar temperature range and was attributed to carrier redistribution among different QDs via the wetting layer. \cite{wang94,yang97,sanguinetti99} Such a mechanism is not possible for droplet-epitaxy QDs due to the lack of a wetting layer and carrier redistribution among different QDs can be safely neglected because the thermal energy is much smaller than the confinement potential of charge carriers in the present experiment, see also \figref{bandStruct}.

We first discuss the physical mechanism governing the decrease in the bright-exciton transition strength qualitatively and return to a more formal discussion later. We first focus on bright excitons because dark excitons do not affect the fast rate. Due to the large size of droplet-epitaxy QDs, excited states with small \OS\ might be thermally activated. In the single-particle picture for example, if the hole populates the first excited state $\ket{2_h}$ (this is more likely because the effective mass of holes is larger than for the electrons) and the electron is in the ground state $\ket{1_e}$, then the dipole transition strength between them is parity forbidden. This is consistent with the fact that we do not see excited states in the PL spectrum. The in-plane symmetry of the QDs implies that there must exist two closely-spaced optically-inactive excited states (they would be degenerate in case of perfect rotational symmetry). As a consequence, a single excitation would be shared between a parity-bright and two parity-dark eigenstates. In the limit $k_BT \gg \Delta E_{HB}$, where $\Delta E_{HB}$ is the energy difference between the ground and excited states, the exciton will populate the bright state with a probability of $1/3$, which means that its effective transition strength will decrease by a factor of three.

\begin{figure}[t!]
	\includegraphics[width=0.4\textwidth]{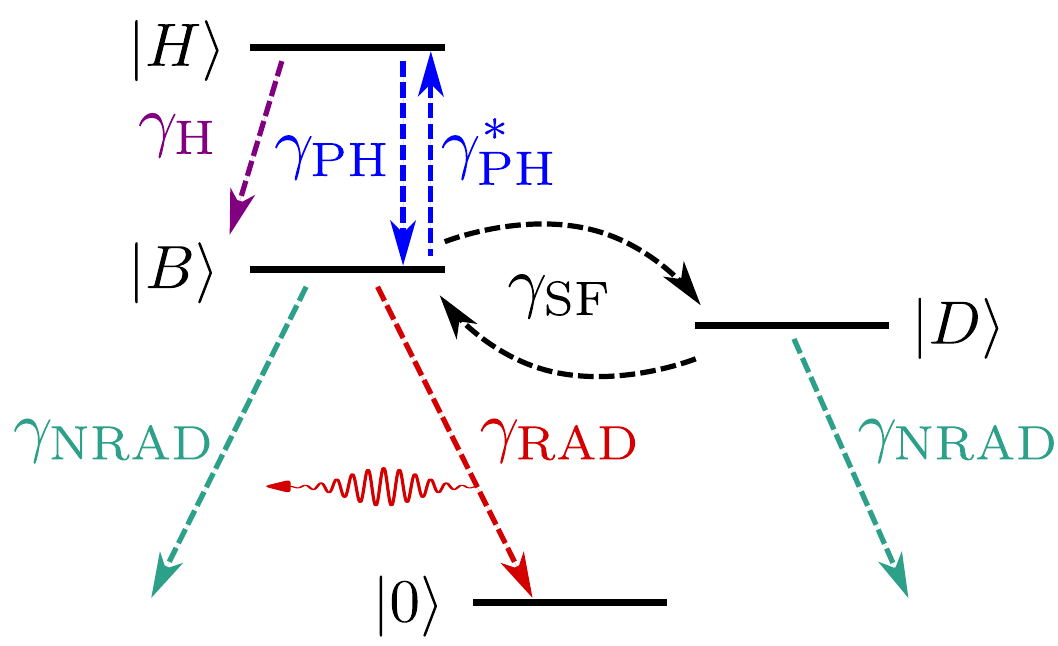}
	\caption{ \label{fig:levelScheme} (Color online) A natural extension of the level scheme presented in \figref{fig:decayDynamics}(a) at elevated temperatures. In large QDs the ground-state bright exciton might be thermally excited to a higher energy (hot) state $\ket{H}$ ($\gamma_\mathrm{PH}^* = \gamma_\mathrm{PH} \times \mathrm{BF}$, where BF is the Boltzmann factor). The hot state can decay back to the bright state via $\gamma_\mathrm{PH}$; aside from that, $\ket{H}$ decays nonradiatively via $\gamma_\mathrm{H}$.}
\end{figure}

We denote the excited eigenstates hot states $\ket{H}$ and the modified level scheme is sketched in \figref{fig:levelScheme}. For simplicity, we first analyze the implications of a single hot state and then we generalize the results.  The hot state decays to the bright state via the phonon-mediated rate $\gamma_\mathrm{PH}$ and becomes populated from the bright state via $\gamma_\mathrm{PH}^*=\gamma_\mathrm{PH}\times \exp\left(-\Delta E_{HB}/k_BT\right)$. Additionally, the hot state decays nonradiatively via $\gamma_\mathrm{H}$. We include only the bright exciton of the hot states in our model to simplify the discussion, however its implications will be addressed for the dark state as well.
Therefore, we set the spin-flip rate $\gamma_\mathrm{SF}=0$ for the moment and solve the rate equations analytically. For the realistic assumption that $\gamma_\mathrm{PH}$ is the fastest rate in the system (phonon scattering is of the order of picoseconds), we obtain for the decay of the bright state
\begin{equation}
\begin{split}
\rho_B(t) &= \frac{1}{1+\mathrm{BF}}\left[ \mathrm{BF}\rho_B(0) - \rho_H(0) \right]e^{-\gamma_\mathrm{PH}(1+\mathrm{BF})t} \\
          &+ \frac{1}{1+\mathrm{BF}}\left[ \rho_B(0) + \rho_H(0) \right]e^{- \frac{\gamma_\mathrm{RAD}+\gamma_\mathrm{NRAD} + \mathrm{BF}\gamma_\mathrm{H}}{1+\mathrm{BF}} t},
\end{split}
\label{eq:BrightHot}
\end{equation}
where $\mathrm{BF}=\exp\left( - \Delta E_{HB}/k_BT \right)$ is the Boltzmann factor. The first term accounts for the build-up of the excitonic population on a phonon scattering time scale, as expected. The population decay is characterized by the second term
\begin{equation}
\begin{split}
\gamma_F &= \frac{\gamma_\mathrm{RAD}+\gamma_\mathrm{NRAD}+e^{- \frac{\Delta E_{HB}}{k_BT} }\gamma_\mathrm{H}}{1+e^{- \frac{\Delta E_{HB}}{k_BT} }}\\
&= \gamma_\mathrm{RAD}^* + \gamma_\mathrm{NRAD}^*.
\end{split}
\label{eq:decayBrightTemp}
\end{equation}
where the asterisk denotes temperature-dependent quantities, and $\gamma_H$ was merged with the temperature-dependent nonradiative rate. It is clear that for thermal energies comparable to $\Delta E_{HB}$, the fast decay rate of the bright state decreases up to a factor of two. In general, the presence of $N$ parity-dark states would decrease the effective transition strength $F$ by a factor of $N+1$, i.e., $F=f\times\gamma_\mathrm{RAD}^{\mathrm{HOM}*}/\gamma_\mathrm{RAD}^\mathrm{HOM}=f/(N+1)$. As a consequence, the almost threefold decrease observed for $F$ (see \figref{fig:dotA_temp}(d)) suggests the presence of two parity-dark states.  At \SI{50}{\kelvin} the fast decay rate is further reduced, which suggests that the effective transition strength continues to decrease and that interaction with even more excited states becomes feasible. The energy difference between the bright and parity-dark states is of the order of several meV, i.e., comparable to the thermal energy in the investigated temperature range. In the present study it is unfortunately not possible to accurately quantify it because parameters such as the energy difference between the hot states, their nonradiative decay rates, etc., are unknown.

It is well known that at elevated temperatures nonradiative decay channels become increasingly important\cite{dai97,mukai97,wu96} and this is indeed what we observe in \figref{fig:dotA_temp}(c), where the nonradiative decay rate increases by about \SI{50}{\percent}. This has a direct impact on the \QE, which diminishes from \SI{70}{\percent} to \SI{40}{\percent}, see \figref{fig:dotA_temp}(d). Interestingly, our data clearly show that it is incorrect to associate a decrease in the fast decay rate with a reduction of nonradiative processes.  

The spin-flip rate in droplet-epitaxy QDs is similar to SK InAs QDs and amounts to several tens of \si{\micro\second^{-1}} at \SI{10}{\kelvin} (cf. Table~\ref{table:decayParameters}). Spin-flip is a phonon-mediated process as discussed in \secref{sec:OSQE}, which implies that it depends on the number of available phonons $N_B$, quantified by the Bose-Einstein distribution
\begin{equation}
N_\mathrm{B} = \frac{1}{\exp(\Delta_\mathrm{BD}/k_BT)-1}.
\end{equation}
In our experiment $k_BT \gg \Delta_\mathrm{BD}$ or $N_\mathrm{B} \gg 1$, which means that the spin-flip rate is given by $\gamma_\mathrm{SF} \approx \gamma_0N_\mathrm{B} \approx k_BT/\Delta_\mathrm{BD}\times \gamma_0$, where $\gamma_0$ is the spin-flip rate at \SI{0}{\kelvin}. By fitting the data with a linear function passing through the origin we obtain a good agreement, as seen in \figref{fig:dotA_temp}(c). We extract a slope of $\gamma_0/\Delta_\mathrm{BD}\approx \SI{15}{\nano\second^{-1}\electronvolt^{-1}}$, and for typical values of $\Delta_\mathrm{BD}\approx \SI{200}{\micro\electronvolt}$\cite{bester03} obtain a zero-temperature spin-flip rate of $\gamma_0 \approx \SI{3}{\micro\second^{-1}}$.

\section{Acoustic-Phonon Broadening and Exciton Size}
\label{sec:LOphonons}

In \secref{sec:OSQE} we have shown that ground-state excitons are strongly confined in droplet-epitaxy QDs. This conclusion was based on the small \OS\ extracted from time-resolved measurements. In this section, we bring further evidence of strong confinement of charge carriers by analyzing the phonon sidebands in the emission spectra of droplet-epitaxy QDs. In particular, we set up a simple fitting routine based on the independent-boson model, and we find that the electron and hole wavefunctions are smaller than the exciton Bohr radius, which defines the limit for strong confinement. The model implemented in this section has been employed to investigate the electron-phonon interaction,\cite{besombes01, peter05,mahan2000} and is used here as a tool to quantify the sizes of the electron and hole wavefunctions.

We consider a two-level system coupled to an acoustic phonon bath. The Hamiltonian of this coupled exciton-phonon system reads

\begin{equation}
	H = E_0c^{\dagger}c + \sum_{\mathbf{k}} \hbar\omega_{\mathbf{k}}\left(b_{\mathbf{k}}^{\dagger}b_{\mathbf{k}} + \frac{1}{2}\right) + c^{\dagger}c\sum_{\mathbf{k}} M_{\mathbf{k}}(b_{\mathbf{k}}^{\dagger}+b_{\mathbf{k}}),
	\label{eq:Hamiltonian}
\end{equation}
where $c^{\dagger}$ and $b_{\mathbf{k}}^{\dagger}$ ($c$ and $b_{\mathbf{k}}$) are the creation (annihilation) operators of the exciton (with energy $E_0$) and the phonon (with momentum $\hbar\mathbf{k}$), respectively. The last term in Eq.~(\ref{eq:Hamiltonian}) denotes the interaction Hamiltonian, where $ M_{\mathbf{k}} $ is the electron-phonon interaction matrix element. The phonon bath represents a continuous set of modes with momentum $\hbar\mathbf{k}$, and each mode has a probability $ M_{\mathbf{k}} $ of interacting with the two-level system. The exciton has a finite lifetime given by its radiative decay rate $\gamma_\mathrm{RAD}$. In order to compute $ M_{\mathbf{k}} $, we follow a number of assumptions: \cite{Besombes2001}

\begin{figure*}
  \centering
  \includegraphics[width=0.9\textwidth]{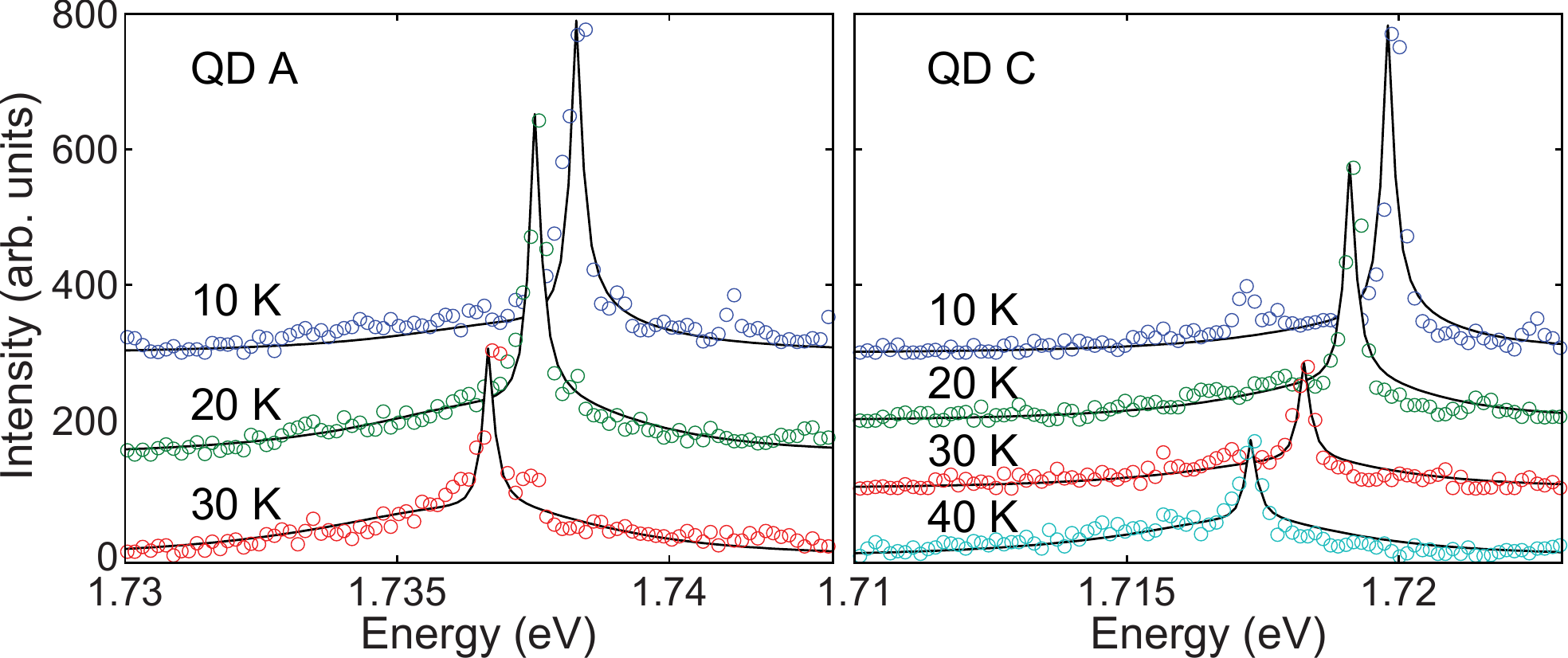}
  \caption{Experimental data (colored circles) along with the fits (solid black lines) for QD A and QD C at different temperatures. All data were recorded under the same conditions with an excitation power density of \SI{286}{\watt\per\square\centi\meter}.}
  \label{fig:fit}
\end{figure*}

(1) The deformation-potential coupling to longitudinal acoustic (LA) phonon modes is the dominant electron-phonon interaction term and, therefore, interactions with transverse acoustic modes and the piezoelectric coupling are neglected.\cite{Takagahara1999}

(2) We consider bulk phonons only, i.e., the QD couples to vibrational modes of the surrounding material, Al$_{0.3}$Ga$_{0.7}$As.

(3) The LA phonon dispersion relation is linear in the relevant energy range $\omega_{\mathbf{k}} = c_s\lvert\mathbf{k}\rvert $, where $c_s$ is the speed of sound in the crystal and is taken to be isotropic (averaged over all directions).

(4) We employ the effective-mass approximation for charge carriers and work in the single-particle picture where electrons and holes are independent entities.

Under these assumptions, the phonon matrix element reads\cite{Besombes2001}
\begin{equation}
	M_{\mathbf{k}} = \mathcal{N}_k\left[D_{e}\bra{\psi_{e}}\mathrm{e}^{i\mathbf{k}\cdot\mathbf{r}}\ket{\psi_{e}} - D_{g}\bra{\psi_{g}}\mathrm{e}^{i\mathbf{k}\cdot\mathbf{r}}\ket{\psi_{g}}\right],
	\label{eq:phonon_matrix_element}
\end{equation}
where $\mathcal{N}_k=\sqrt{\hbar\lvert\mathbf{k}\rvert /2dc_sV}$, $\psi_{e}$ ($\psi_g$) is the slowly-varying envelope function of the electron (hole), and $V$ is the quantization volume. The following constants are used for Al$_{0.3}$Ga$_{0.7}$As: the density $d= \SI{4805}{\kilo\gram\per\cubic\meter}$,  the speed of sound $c_s = \SI{5396}{\meter\per\second}$,\cite{Adachi1993} and the deformation potentials $D_g = \SI{5.6}{\electronvolt}$ and $D_e = \SI{-11.5}{\electronvolt}$.\cite{Peter2004} We assume lens-shaped wavefunctions with a Gaussian spatial profile
\begin{equation}
	\psi_\nu(\mathbf{r})=\frac{1}{\pi^{3/4}\sigma_{\nu,\rho} \sqrt{\sigma_{\nu,z}}} \mathrm{e}^{-\frac{\rho^2}{2\sigma_{\nu,\rho}^2}} \mathrm{e}^{-\frac{z^2}{2\sigma_{\nu,z}^2}},
\end{equation}
where $\rho=\sqrt{x^2+y^2}$ is the in-plane radial coordinate, and $\sigma$ is the half-width at half maximum (HWHM).
The matrix element is evaluated to be
\begin{equation}
	M_{\mathbf{k}} = \mathcal{N}_k\left[ D_e \mathrm{e}^{-\frac{1}{4}(\sigma_{e,\rho}^2k_\rho^2+\sigma_{e,z}^2k_z^2)}- D_g \mathrm{e}^{-\frac{1}{4}(\sigma_{g,\rho}^2k_\rho^2+\sigma_{g,z}^2k_z^2)} \right].
\end{equation}
The phonon contribution function $\Phi(t)$ is derived directly from the interaction Hamiltonian in Eq.~(\ref{eq:Hamiltonian}) and is defined as\cite{mahan2000}
\begin{equation}
	\Phi(t) = \sum_{\mathbf{k}}\frac{\lvert M_{\mathbf{k}}\rvert^2}{(\hbar \omega_k)^2}\left[ i\sin(\omega_\mathbf{k}t)+ \left[1-\cos(\omega_\mathbf{k}t)\right](2n_\mathbf{k}+1)\right],	
	\label{eq:Phi}
\end{equation}
where $ n_{\mathbf{k}} =(\mathrm{e}^{\hbar\omega_{\mathbf{k}}/k_BT}-1)^{-1} $ is the thermal occupation function. For a large quantization volume, we can transform the sum over $\mathbf{k}$ into an integral as $ \sum_{\mathbf{k}} \to \frac{V}{(2\pi)^3}\int \mathrm{d}\mathbf{k} $, so that Eq.~(\ref{eq:Phi}) becomes 
\begin{equation}
\begin{split}
	\Phi(t) &= C\int\limits_0^{\infty}k\mathrm{d}k \int\limits_{0}^{1}\mathrm{d}y \left\lvert	D_e \mathrm{e}^{-\frac{\sigma_{e}^2}{4}k^2(1-\xi_e y^2)}- D_g\mathrm{e}^{-\frac{\sigma_{g}^2}{4}k^2(1-\xi_g y^2)} \right\rvert^2 
	\\
	&\times \left[ i\sin(\omega_\mathbf{k}t)+ \left[1-\cos(\omega_\mathbf{k}t)\right](2n_\mathbf{k}+1)\right],
	\end{split}
\end{equation}
where $C = 1/4\pi^2 d c_s^3\hbar$ and $\xi_\nu = 1-\sigma_{\nu,z}^2/\sigma_{\nu,\rho}^2$.
The integration over $y$ is performed analytically, and the integral over $k$ is evaluated numerically. Finally, the emission spectrum is computed in the following way \cite{mahan2000}

\begin{equation}
	S(\omega) = \int\limits_{-\infty}^{+\infty}\mathrm{d}t \mathrm{e}^{-i(\omega-\omega_0-i\gamma_{\mathrm{RAD}}/2)t} \mathrm{e}^{-\Phi(t)},
	\label{eq:IBM_S_omega}
\end{equation}
where $\omega_0$ is the emission frequency of the QD. 

\begin{table}
\centering
  \begin{tabular}{c c c c}\botrule
 &  $ \sigma_{e,\rho} $ (nm) & $ \sigma_{g,\rho} $ (nm) & $f$\\
    \colrule
QD A & 2.4  & 2.4  & 16.6 \\
QD C & 3.6  & 1.9  & 9.4 \\
 \botrule
  \end{tabular}
  \caption{Fitted sizes (HWHM) of the electron and hole wavefunctions and the resulting oscillator strength $f$.}
\label{table:fit_values}
\end{table}

We fit the experimental data using a least-square approach so that the sum of the squared residuals is minimized. Following the observations from AFM measurements, we fix the ratio between the wavefunction height and radius $\sigma_{\nu,\rho} = \alpha\sigma_{\nu,z}$ with $\alpha=3$. As a consequence, we have only two independent fitting parameters, namely the size of the hole and electron wavefunctions. For QD C, the spectrum is fitted at the highest recorded temperature (\SI{40}{\kelvin}) because the signal coming from the phonon sidebands increases with temperature, thus enhancing the accuracy of the fitted parameters. For QD A, the fitting is performed at \SI{30}{\kelvin} because at higher temperatures there is an additional line appearing in the spectrum, see \figref{fig:dotA_temp}(a), which renders the fit difficult to realize. For the data at lower temperatures we do not fit but simply plot the evaluated emission spectrum. 

Figure~\ref{fig:fit} shows the spectra of QDs A and C with very good agreement between theory and experiment. The extracted sizes (HWHM of 2--\SI{4}{\nano\meter}) are well below the exciton Bohr radius (\SI{11.2}{\nano\meter}). Thus, this independent analysis confirms the conclusions drawn from time-resolved measurements, namely that ground-state excitons are strongly confined in droplet-epitaxy QDs. We can give an estimate of the oscillator strength using \eqref{eq:OSstrongconf}, which agrees reasonably well with experiment, compare Tables~\ref{table:decayParameters} and \ref{table:fit_values}.

\section{Conclusion}
In this paper, we have presented an extensive study of the optical properties and decay dynamics of large strain-free droplet-epitaxy GaAs QDs. From the measurements, we draw several important conclusions:

(1) The droplet-epitaxy QDs exhibit an \OS\ and \QE\ of about 9 and \SI{75}{\percent}, respectively.

(2) Ground-state excitons are strongly confined despite the large size of the droplet-epitaxy QDs observed in AFM measurements. This is caused by material interdiffusion occurring between the QDs and the surrounding matrix, which creates a localized potential minimum that traps carriers in a region of space smaller than the exciton Bohr radius. This physical picture is supported by two independent analyses: the \OS\ extracted from time-resolved measurements and the sizes of electron and hole wavefunctions evaluated by fitting the spectral phonon sidebands.

(3) For some QDs, the bright exciton is thermally activated to parity-dark eigenstates with temperature. As a consequence, the radiative lifetime of bright excitons is substantially prolonged and the effective transition strength decreases from 10 to 4 as the temperature is raised from \SI{10}{\kelvin} to \SI{40}{\kelvin}. Additionally, the nonradiative recombination rate is increased by almost a factor of two in the same temperature range. Both affect the \QE, which attains a value of only \SI{40}{\percent} at \SI{40}{\kelvin}.

Our findings show that droplet-epitaxy GaAs QDs, similarly to the commonly used InAs QDs, exhibit non-negligible nonradiative processes. This is likely due to the low-temperature growth of the QDs and of the capping layer forming a crystalline structure of low quality, which is not fully restored by thermal annealing. Although we have not found a giant \OS\ in these QDs, we believe that better growth techniques have the capability of improving this aspect owing to the lack of strain in these structures.

Finally, we mention that the general conclusion that the actual exciton size can be significantly smaller than the QD size has also been reached for other material systems. By analyzing phonon-broadened spectra, Rol et al.~\cite{rol07} found that the excitons confined in GaN/AlN QDs are much smaller than the spatial extent of the QD. Stobbe et al.~\cite{stobbe10} extracted a small \OS\ of large InGaAs QDs of about 10 by controllably modifying the LDOS at the position of the emitter. The latter work points to the same physical situation, namely that the induced material inhomogeneities during growth create a non-uniform potential profile, which strongly confines excitons. Therefore, engineering large QDs with large excitons and giant oscillator strength represents a future challenge in the field of nanophotonics.

\section*{Acknowledgements}
We gratefully acknowledge the financial support from the Danish council for independent research (natural sciences and technology and production sciences), and the European Research Council (ERC consolidator grant "ALLQUANTUM"). The KIST researchers acknowledge the support from the KIST institutional program (including the dream project), the grant 2012K001280 and the GRL Program through MEST.

\bibliographystyle{nature}
\bibliography{bibliography}

%\printbibliography

\end{document}